%% file: main_combined.tex
\documentclass[twoside]{article}

\usepackage{aistats2020}
\usepackage[round]{natbib}
\usepackage{amsmath,xfrac,stackengine,amsthm,amssymb}
\usepackage{hyperref}
\usepackage{url,xcolor}
\usepackage{multirow}
\usepackage{subfig,float}
\usepackage{algorithm} 
\usepackage[noend]{algpseudocode}
\input{math_commands.tex}

\usepackage{caption}
\renewcommand{\paragraph}[1]{\vspace{2mm}\noindent\textbf{#1}}
\newcommand{\diag}[1]{\text{diag}(#1)}
\newcommand{\trace}[1]{\text{tr}(#1)}

\newcommand{\norm}[1]{\left\lVert#1\right\rVert}

\newcommand{\G}{\mathcal{G}}  
  
\renewcommand{\P}{\mathcal{P}}  
\newcommand{\V}{\mathcal{V}}  
\renewcommand{\E}{\mathcal{E}}  
\renewcommand{\S}{\mathcal{S}}  

\newcommand{\I}{\bm{I}}
\newcommand{\C}{\bm{C}}  
\newcommand{\mPi}{\bm{\Pi}} 

\newcommand{\cL}{\bm{L}} 
\newcommand{\nrL}{\bm{\mathcal{L}}} 
\newcommand{\signL}{\tilde{\bm{\mathcal{L}}}} 
\newcommand{\rwL}{\nrL^{\textit{rw}}} 
\newcommand{\rwu}{\vu^{\textit{rw}}}

\newcommand{\SD}{\textit{SD}}

\theoremstyle{style}
\newtheorem{theorem}{Theorem}[section]
\newtheorem{proposition}{Proposition}[section]
\newtheorem{definition}[theorem]{Definition}
\newtheorem{property}[theorem]{Property}
\newtheorem{corollary}[theorem]{Corollary}

\begin{document}

%

%

\twocolumn[

\aistatstitle{Graph Coarsening with Preserved Spectral Properties}


\aistatsauthor{ Yu Jin \And Andreas Loukas \And  Joseph F. JaJa }

\aistatsaddress{ University of Maryland \And  EPFL \And University of Maryland } 
]

\begin{abstract}
  Large-scale graphs are widely used to represent object relationships in many real world applications. The occurrence of large-scale graphs presents significant computational challenges to process, analyze, and extract information. Graph coarsening techniques are commonly used to reduce the computational load while attempting to maintain the basic structural properties of the original graph. As there is no consensus on the specific graph properties preserved by coarse graphs, how to measure the differences between original and coarse graphs remains a key challenge. In this work, we introduce a new perspective regarding the graph coarsening based on concepts from spectral graph theory. We propose and justify new distance functions that characterize the differences between original and coarse graphs. We show that the proposed spectral distance naturally captures the structural differences in the graph coarsening process. In addition, we provide efficient graph coarsening algorithms to generate graphs which provably preserve the spectral properties from original graphs. Experiments show that our proposed algorithms consistently achieve better results compared to previous graph coarsening methods on graph classification and block recovery tasks. 
\end{abstract}

\section{INTRODUCTION}

Graphs are widely used to represent object relationships in real-world applications. As many applications involve large-scale graphs with complex structures, it is generally hard to explore and analyze the key properties directly from large graphs. Hence the graph coarsening techniques have been commonly used to facilitate the process \citep{liu2016graph, chevalier2009comparison}.

Generally speaking, the aim of any graph reduction scheme is to reduce the number of nodes and edges of a graph, while also ensuring that the ``essential properties'' of the original graph are preserved. 
The question of what these properties should be remains inconclusive, but there is significant evidence that they should relate to the spectrum of a graph operator, such as the adjacency or normalized Laplacian matrix. 
A long list of theorems in spectral graph theory show that the combinatorial properties of a graph are aptly captured by its spectrum. As such, graphs with similar spectrum are generally regarded to share similar global and local structure~\citep{van2003graphs, banerjee2008spectrum}.
Based on this realization, modern graph sparsification techniques~\citep{spielman2011graph, jovanovic2012spectral, batson2013spectral} have moved on from previously considered objectives, such as cut and shortest-path distance preservation, and now aim to find sparse spectrally similar graphs.

In contrast to graph sparsification, in coarsening there has been little progress on attaining spectrum preservation guarantees. The foremost roadblock seem to lie in defining what spectral similarity should entail for graph of different sizes. The original and coarse graphs now have different number of eigenvalues and eigenvectors which prohibits a direct comparison. To circumvent this issue, recent works have considered restricting the guarantees to a subset of the spectrum~\citep{loukas2018spectrally, loukas2019graph}. Focusing however only on the first few eigenvalues and eigenvectors also means that important information of the graph spectrum is ignored.

In this work, we start by reconsidering the fundamental spectral distance metric~\citep{jovanovic2012spectral,gu2015spectral,jovanovic2015some,jovanovic2014spectral}, which compares two graphs by means of a norm of their eigenvalue differences. This metric is seemingly inappropriate as it necessitates two graphs to have the same number of eigenvalues. However, we find that in the context of coarsening this difficulty is easily circumvented by substituting the coarse graph with its lifted counterpart: the latter contains exactly the same information as the former while also having the correct number of eigenvalues. 

Our analysis shows that the proposed distance naturally captures the graph changes in the graph coarsening process. In particular, when the graph coarsening merges nodes that have similar connections to the rest of the graph, the spectral distance is provably small. By merging similarly connected nodes, nodes and edges in coarse graphs are able to represent the connectivity patterns of the original graphs, thus preserving structural and connectivity information. Following the graph coarsening framework and the new notion of spectral distance, we provide two efficient graph coarsening algorithms to generate coarse graphs which provably  preserve the spectral properties. 

Our contributions are summarized as follows:
\begin{itemize}
	\item We show how the spectral distance~\citep{jovanovic2012spectral,gu2015spectral,jovanovic2015some,jovanovic2014spectral}, though originally restricted to graphs of the same size, can be utilized to measure how similar a graph is with its coarsened counterpart. 
	\item We provide a rigorous justification that the new spectral distance accurately captures the graph structural changes in the graph coarsening process.
	\item We present efficient algorithms to generate coarse graphs which provably preserve spectral properties from original graphs.
	\item In the experiment, we show that the proposed methods outperform other graph coarsening algorithms on several graph related tasks.
\end{itemize}
\section{RELATED WORK}
\label{sec:related-work}
Recent work have proposed to coarsen graphs by preserving the spectral properties of the matrix representations of graphs \citep{loukas2018spectrally, loukas2019graph, durfee2019fully, purohit2014fast, hermsdorff2019unifying}. 
%
For example, \citet{loukas2019graph} proposed to preserve the action of the graph Laplacian with respect to an (eigen)-space of fixed dimension, arguing that this suffices to capture the global properties of graph relevant to partitioning and spectral clustering. \citet{durfee2019fully} aimed to preserve the all-pairs effective resistance. \citet{garg2019solving} defined a cost based on the theory of optimal transport. Saket et al. suggested a Minimum Description Length (MDL) principle relevant to unweighted graphs \citep{navlakha2008graph}. Most of these distance functions are specific to special applications, how to define an application-independent graph coarsening framework remains a challenge.

There is a vast literature on characterize graphs with graph spectral properties \citep{jovanovic2012spectral, tsitsulin2018netlsd, dong2019network}. Previous work defined distance functions based on Laplacian eigenvalues which measure differences between graphs \citep{jovanovic2012spectral, gu2015spectral}. Spielman and Teng introduced a new notion of spectral similarity for two graphs in their graph sparsification framework  \citep{batson2013spectral,spielman2011spectral}. Recently, Tsitsulin et al. proposed an efficient graph feature extractor, based on Laplacian spectrum, for comparisons of
large graphs \citep{tsitsulin2018netlsd}. Dong uses spectral densities to visualize and estimate meaningful information about graph structures \citep{dong2019network}. Although the graph spectrum have been widely used in many graph applications, it still remains little explored in the context of graph coarsening. 
\section{PRELIMINARIES}
Let $\G = (\V, \E, \mW)$ be a graph, with $\V$ a set of $N = |\V|$ nodes, $\E$ a set of $M = |\E|$ edges, and $\mW \in \mathbb{R}^{N\times N}$ the adjacent matrix. We denote by $v_i$ as the node indexed at $i$, $\vw(i) \in \mathbb{R}^N$ as the vector of all possible edge weights associated with $v_i$ and $d(i) = \sum_{j = 1}^N \mW(i, j)$ as the node degree. Graphs defined in this work are weighted, undirected, and possess no isolated nodes (i.e. $d(i) > 0$ for all $v_i$).
Furthermore, the combinatorial and normalized Laplacians of $\G$ are defined as
\begin{equation}
\cL = \mD^{-1} \mW \quad \text{and} \quad \nrL = \mI_N - \mD^{-1/2}\mW\mD^{-1/2}
\end{equation}
respectively, where $\mI_N$ is the $N$-dimensional identity matrix and $\mD$ is the diagonal degree matrix with $\mD(i,i) = d(i)$.

\begin{figure*}[t!]
	\centering
	\begin{minipage}{.9\columnwidth}
		\centering
		\includegraphics[width=\linewidth]{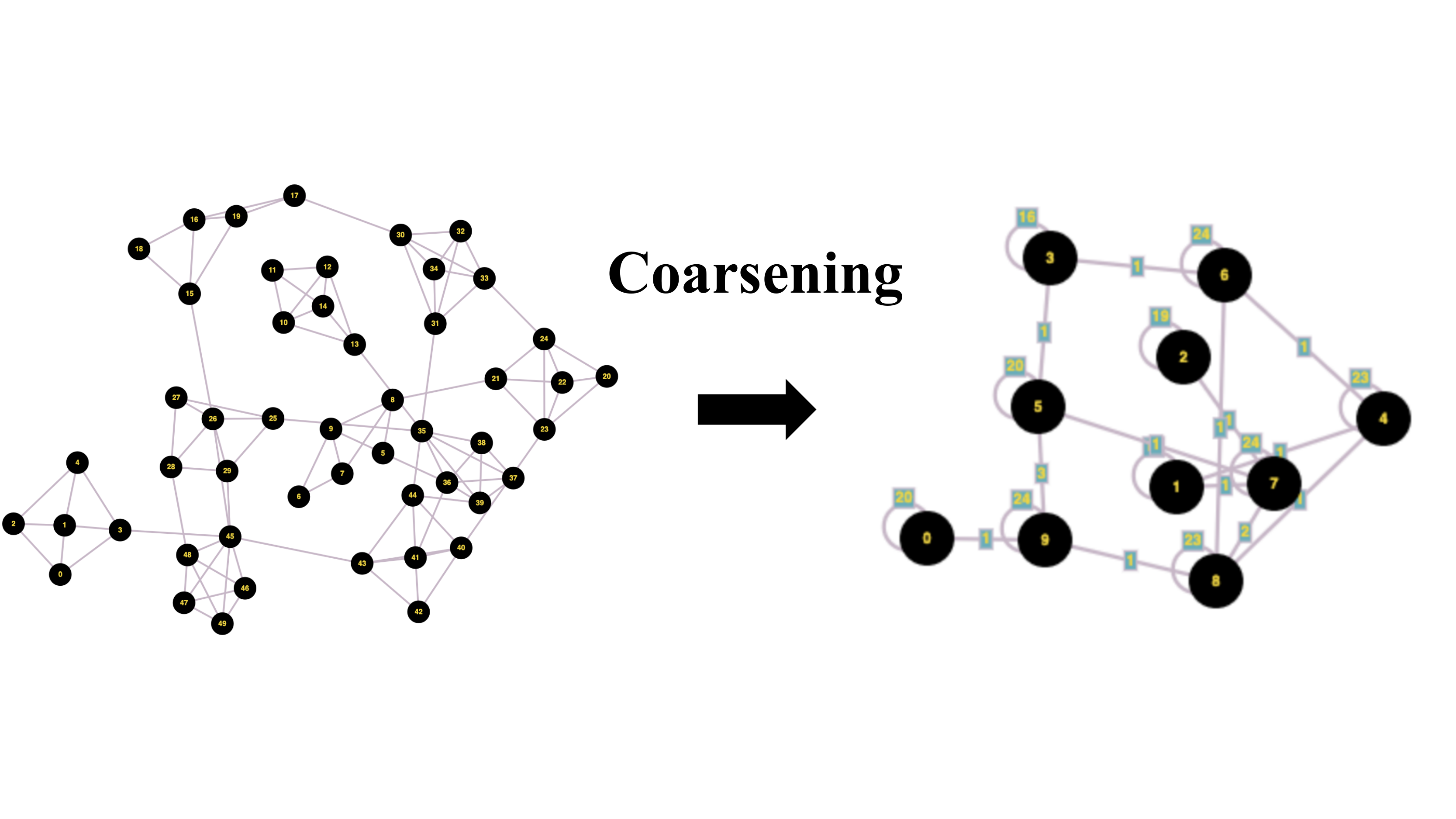}
	\end{minipage}
	\begin{minipage}{.49\columnwidth}
		\centering
		\includegraphics[width=\linewidth]{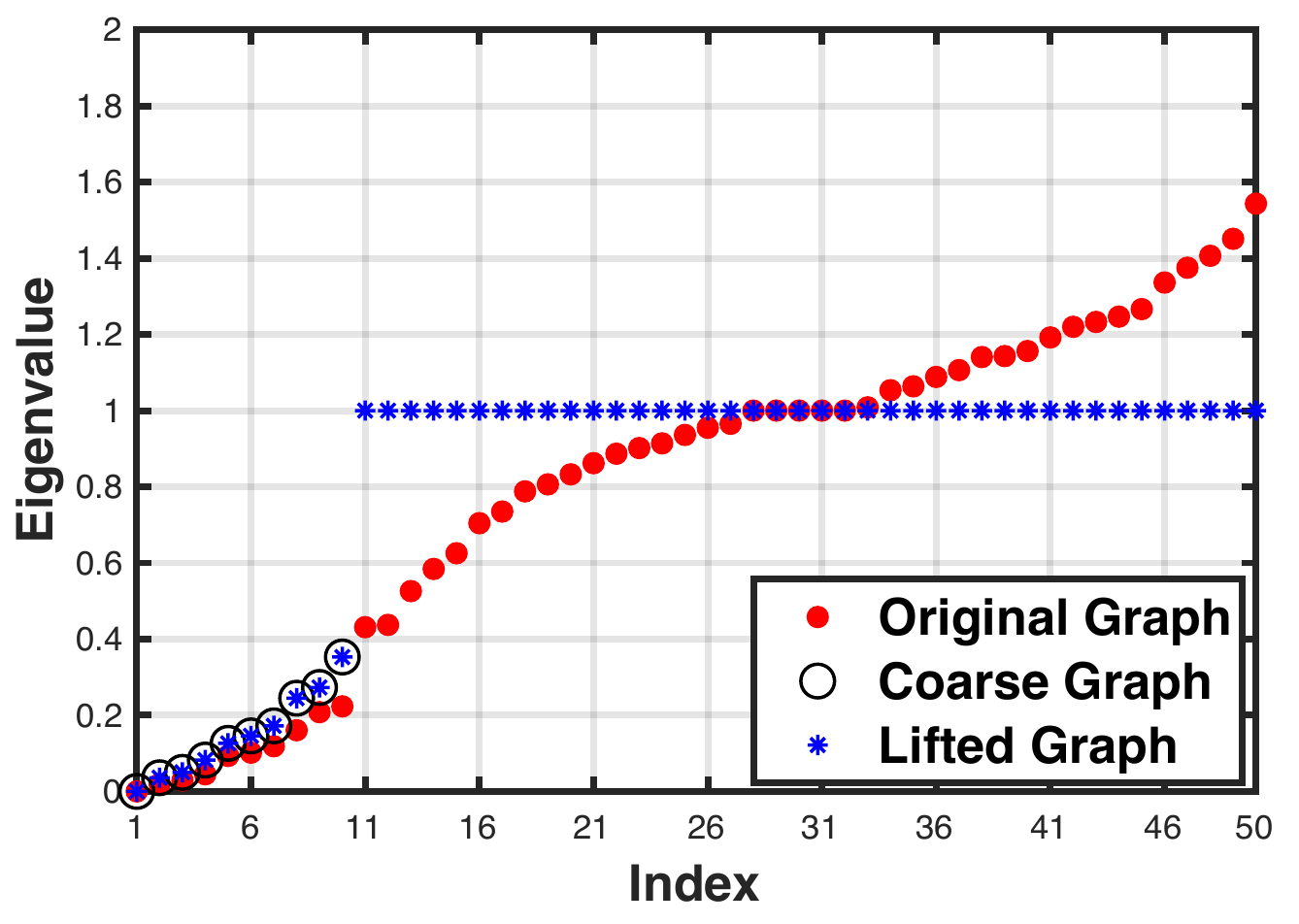}
	\end{minipage}
	\begin{minipage}{.49\columnwidth}
		\centering
		\includegraphics[width=\linewidth]{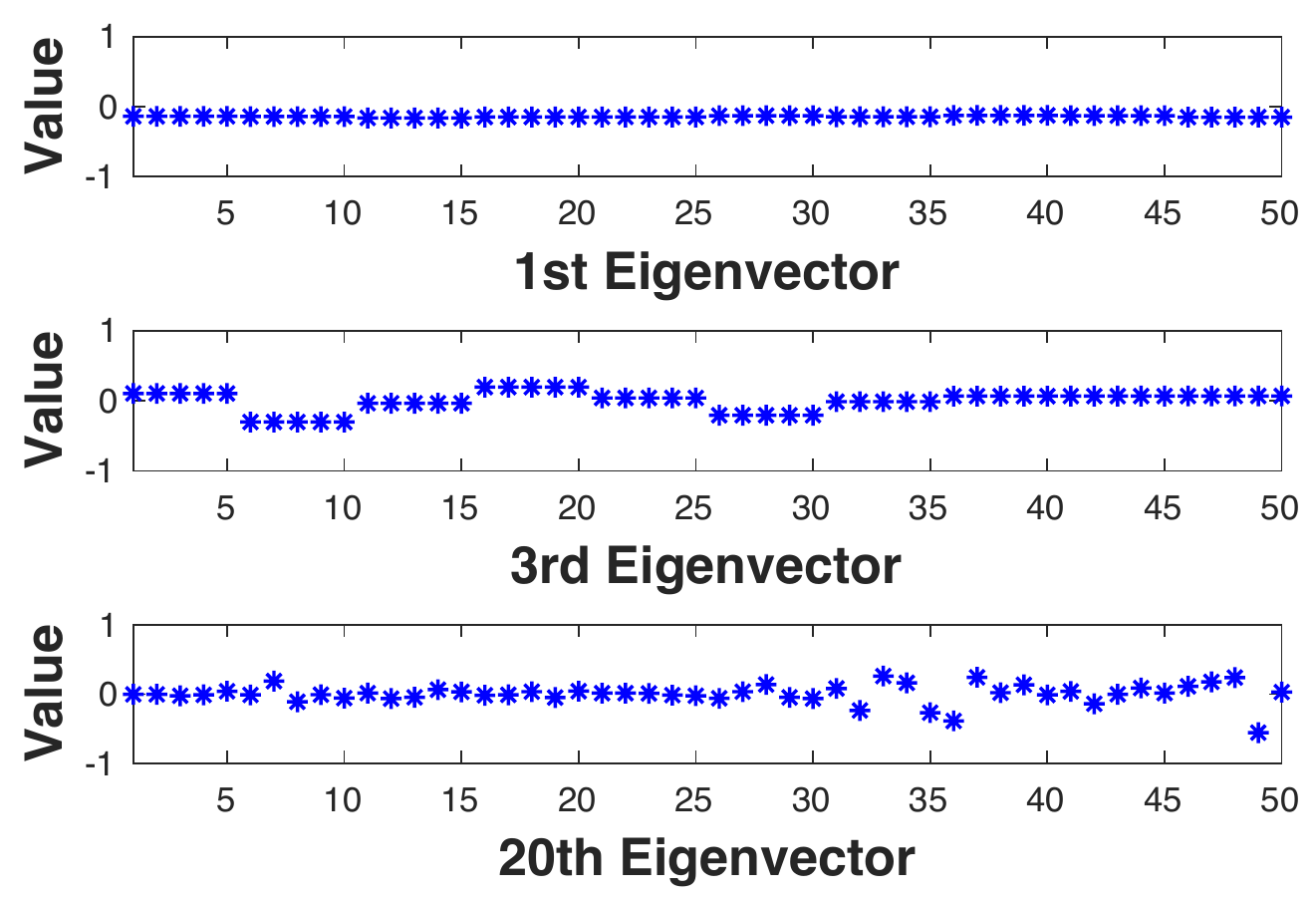}
	\end{minipage}
	\caption{\textbf{Left}: an example illustrating the graph coarsening process. The original graph is a random graph sampled from stochastic block model with 50-node and 10 predefined blocks. \textbf{Right}: Eigenvalues and eigenvectors of normalized Laplacian matrices of original, coarse and lifted graphs. The eigenvalues of coarse graphs align with the eigenvalues of original graphs. In addition, the eigenvectors indicate the block membership information.}
	\label{fig:spectral_example}
\end{figure*}

\subsection{Graph Coarsening} 
The coarse graph $\G_c = (\V_c, \E_c, \mW_c)$  with $n = |\V_c|$ is coarsened from the original graph $\G$ with respect to a set of non-overlapping graph partitions $\P = \{\S_1, \S_2, \ldots, \S_n \} \subset \V$. Each partition $\S_p$ corresponds to a ``supernode" denoted by $s_p$ and the ``superedge" connecting the supernodes $\mW_c(p, q)$ has the edge weight as the accumulative edge weights between nodes in the corresponding graph partitions $\S_p$ and  $\S_q$:
$$
\mW_c(p, q) = w(\S_p, \S_q):= \sum_{v_i \in \S_p, v_j \in \S_q} \mW(i, j)
$$
Let $\mP \in \mathbb{R}^{n \times N}$ be the matrix whose columns are partition indicator vectors,
$$
\label{eqn:P}
\mP(p, i) = \begin{cases}
1, & \text{if $v_i \in \S_p$} \\
0, & \text{othewise}.
\end{cases}
$$
It is then well known that the weight matrix $\mW_c$ of the coarse graph $\G_c$ abides to
$$
\mW_c = \mP\mW\mP^\top. 
$$

The definition of the coarsened Laplacian matrices follows directly:
$$
\cL_c = \mD_c^{-1}\mW_c \quad \text{and} \quad \nrL_c = \I_n - \mD_c^{-1/2}\mW_c\mD_c^{-1/2}.
$$
Similarly to the adjacency matrix, the combinatorial Laplacian of the coarse graph can be obtained by multiplying $\cL$ with the $\mP$ matrix as  
$
\cL_c = \mP \cL \mP^\top.
$
The same however doesn't hold for the normalized Laplacian since $\mP \nrL \mP ^\top \neq \nrL_c$.

\subsection{Graph Lifting} 
\label{sec:lift}
We define $\G_l = (\V, \E_l, \mW_l)$ to be the graph lifted from the coarse graph $\G_c$ with respect to a set of non-overlapping partitions $\P$. In graph lifting, each node $s_p$ of the coarse graph is lifted to $|\S_p|$ nodes and nodes in the lifted graph are connected by edges with weight as the edge weight normalized by the size of partitions. Specifically, for any $v_i \in \S_p$ and $v_j \in \S_q$ we have 
\begin{align}
    \label{eqn:lift}
    \mW_l(i,j) & = \frac{w(\S_p, \S_q)}{|\S_p| |\S_q|} =  \frac{\sum_{v_i' \in \S_p, v_j' \in S_q} \mW(i',j')}{|S_p| |S_q|} \nonumber\\ 
    & = \frac{\mW_{c}(p,q)}{|S_p| |S_q|}
\end{align}
When $\S_p = \S_q = \S$, the weight $\mW_l(i,j)$ can be seen to be equal to the weight of all edges in the subgraph induced by $\S$ normalized by $|\S|^2$. It easily follows that if $\mW(i,j)$ is the same for every $v_i,v_j \in \S$, then also $\mW_l(i,j) = \mW(i,j)$, i.e., in-partition weights are exactly preserved by successive coarsening and lifting. 
The above combinatorial definition possesses can be expressed in an algebraic form in terms of the the pseudo-inverse $\mP^+$ of $\mP$, (i.e., $\mP \mP^+ = \I$), whose elements are given by:
$$
\mP^+(j, p) = 
\begin{cases} 
\frac{1}{|\S_p|} & \text{if } v_j \in \S_p \\
0       & \text{otherwise}. 
\end{cases}
$$
With this in place, the adjacency matrices of the lifted and coarse graphs are connected by the following relations: 
$$
\mW_l = \mP^+ \mW_c \mP^\mp \quad \text{ and } \quad \mW_c = \mP \mW_l \mP^\top.
$$
The following equation reveals that lifting preserves the connectivity up to a projection onto the partitions:
\begin{align}
\label{eqn:pi}
    \mW_l = \mP^+ \mW_c \mP^\mp = \mP^+ \mP \mW \mP^\top \mP^\mp = \mPi \mW \mPi^\top = \mPi \mW \mPi,
\end{align}
where $\mPi = \mP^+\mP$ is a projection matrix, with $\mPi\mPi = \mP^+\mP\mP^+\mP = \mP^+\mP = \mPi$. 

The lifted Laplacian matrices are given by
\begin{align}
\cL_l = \mP^+\cL_c\mP^\mp \quad \text{and} \quad \nrL_l = \C^\top \nrL_c \C,
\end{align}
where $\C \in \mathbb{R}^{n\times N}$ is the \textit{normalized coarsening matrix} whose entries are given by: 
$$
\label{eqn:cluster}
\C(p, i) = 
\begin{cases}
\frac{1}{\sqrt{|\S_p|}} & \text{if } v_i \in \S_p \\
0 			 & \text{otherwise,}
\end{cases}
$$
such that $\C^\top = \C^+$ and $\C^\top\C = \mP^\mp\mP = \mPi$.
In this manner, we have 
\begin{align}
   \label{eqn:cl}
   \cL_c = \mP \cL_l \mP^\top \quad \text{and} \quad \nrL_c = \C \nrL_l \C^\top. 
\end{align}




\section{SPECTRAL DISTANCE}

We start by briefly reviewing some basic facts about the spectrum associated with the Laplacian matrix of a coarse graph. We then demonstrate how to exploit these properties in order to render the classical spectral distance metric amenable to (coarse) graphs of different sizes.


\subsection{The Coarse Laplacian Spectrum} 

Denote the eigenvalues and eigenvectors of the normalized Laplacian matrices as $\bm{\lambda}$ and $\vu$, respectively, with $\nrL = \mU\bm{\Lambda} \mU^\top$ where the $i$-th column of $\mU$ corresponds to $\vu_i$ and $\bm{\Lambda} = \diag{\bm{\lambda}}$. The eigenvalues are ordered in the non-decreasing order. 

\begin{property}[\textbf{Interlacing}. Section 5.3 in \citep{butler2008eigenvalues}] 
	\label{property:interlace}
	The normalized Laplacian eigenvalues of the original and coarsened graphs satisfy 
	$$
	\bm{\lambda}(i) \le \bm{\lambda}_c(i) \le \bm{\lambda}(i+N-n) \quad \text{for all} \quad i = 1, \ldots, n.
	$$
\end{property}

Property \ref{property:interlace} is a general interlacing inequality capturing pairwise difference between original and coarse graphs \citep{chung1997spectral,butler2007interlacing}. The equality holds under different settings of graph structures and the coarsening matrices.

\begin{property}[\textbf{Eigenvalue Preservation}]
	\label{property:eigenvalue} 
	 The normalized Laplacian eigenvalues of the lifted graph contain all eigenvalues of  the coarse graph and eigenvalue $1$ with $(N-n)$ multiplicities.
\end{property}


	

\begin{property}[\textbf{Eigenvector Preservation}]
	\label{property:eigenvector}
	The eigenvectors of the coarse graph lifted by $\mC$, i.e. $\vu_l = \mC\vu_c$ are the eigenvectors of $\nrL_l$.
\end{property}

Property \ref{property:eigenvalue} and \ref{property:eigenvector} state that lifted graphs preserve most spectral properties of the coarse graph. Thus, they can be used as a proxy to define the distance function \citep{toivonen2011compression}. Figure \ref{fig:spectral_example} shows an example illustrating the graph coarsening process and spectral properties.

\subsection{Spectral Distance} 
\label{sec:sd}
Two notions of the spectral distance are proposed to quantify the difference between original and coarse graphs. We first use the lifted graph as the ``proxy" of the coarse graph to define the \emph{full spectral distance}.

\begin{definition}
	The full spectral distance between graph $\G$ and $\G_c$ is defined as follows: 
	$$
	\label{eqn:full}
	\SD_{\text{full}}(\G, \G_c) = \norm{\mathbf{\bm{\lambda}} - \bm{\lambda}_l }_1 = \sum_{i = 1}^{N}|\bm{\lambda}(i) - \bm{\lambda}_l(i)|,
	$$
	where $\bm{\lambda}$ and $\bm{\lambda}_l$ are the eigenvalues of the original and lifted graphs.
\end{definition}
As the original and lifted graphs have the same number of nodes, we directly use vector distance metric to measure the pairwise differences between eigenvalues. However, the definition requires computing all eigenvalues of original graphs regardless of the coarse graph size, which is computationally expensive especially for large graphs. 

This motivates us to define the \emph{partial spectral distance} by selecting part of the terms in the full spectral distance definition.
We expand the full spectral distance into three terms as follows with $k_1$ and $k_2$ defined as $k_1 = \arg\max_i \{i: \bm{\lambda}_c(i) < 1\}, k_2 = N-n+k_1$,
\begin{align}
\SD_{\text{full}}(\G, \G_c)  & = \sum_{i = 1}^{N}|\bm{\lambda}(i) - \bm{\lambda}_l(i)| \nonumber \\
& = \sum_{i = 1}^{k_1}|\bm{\lambda}(i) - \bm{\lambda}_l(i)| \\
& + \sum_{i = k_1+1}^{k_2}|\bm{\lambda}(i) - \bm{\lambda}_l(i)| + \sum_{i = k_2+1}^{N}|\bm{\lambda}(i) - \bm{\lambda}_l(i)|  \nonumber \\
& = \sum_{i = 1}^{k_1}|\bm{\lambda}(i) - \bm{\lambda}_c(i)| \\
& + \sum_{i = k_1+1}^{k_2}|\bm{\lambda}(i) - 1| + \sum_{i = k_2+1}^{N}|\bm{\lambda}(i) - \bm{\lambda}_c(i-n+N)| \nonumber 
\end{align}
The last equation is from the Property \ref{property:eigenvalue} where $\mathbf{\bm{\lambda}}_l$ contains eigenvalues of the coarse graph as well as eigenvalue 1 with $N-n$ multiplicities. The eigenvalue $\bm{\lambda}_l$ satisfies 
$$
\bm{\lambda}_l(i) = 
\begin{cases}
\bm{\lambda}_c(i) & i \le k_1 \\
1 & k_1 +1 \le i \le k_2 \\
\bm{\lambda}_c(i-N+n) & i > k_2\\
\end{cases}
$$ 
We define the \emph{partial spectral distance} as the full spectral distance excluding the terms where $\bm{\lambda}_l = 1$, that is,
\begin{definition}
	\label{def:partial}
	The partial spectral distance between graph $\G$ and $\G_c$ is defined as 
	$$
	\SD_{\text{part}}(\G, \G_c) = \sum_{i = 1}^{k}|\bm{\lambda}(i) - \bm{\lambda}_c(i)|  + \sum_{i = k+1}^{n}|\bm{\lambda}_c(i) - \bm{\lambda}(i+N-n)|
	$$
	where $k = \arg\max_i \{i: \bm{\lambda}_c(i) < 1\}$.
\end{definition}

For the partial spectral distance, we only need to compute $n$ rather than $N$ eigenvalues of the normalized Laplacian of the original graph, which significantly reduces the computational cost.

The full and partial spectral distances are related by,
$$
\SD_{\text{full}}(\G, \G_c) = \SD_{\text{part}}(\G, \G_c) + \sum_{i = k_1+1}^{k_2}|\bm{\lambda}(i) - 1|
$$
The excluded terms $\sum_{i = k_1+1}^{k_2}|\bm{\lambda}(i) - 1|$ measure the closeness of the original Laplacian eigenvalues and eigenvalue $1$. The two definitions are equivalent when the normalized Laplacian of the original graph $\nrL_N$ contains $N-n$ eigenvalue 1. The condition is equivalent to when the adjacent matrix $W$ is \emph{singular} with $N-n$ algebraic multiplicity of the eigenvalue 0 \citep{sciriha2007characterization, al2018singular}. In the graph coarsening framework, when the merged nodes have similar connections, the adjacent matrix has eigenvalues close to 0.  As a result, the excluded terms are close to 0. Thus, the full and partial spectral distance are close for tackling the graph coarsening problems. 

Note that both definitions of spectral distance are proper distance metrics over the space of graph Laplacian eigenvalues. However, the spectral distance are not able to distinguish  graphs with the same sets of Laplacian eigenvalues (cospectral graphs \citep{van2003graphs}). Thus there could exist multiple coarse graphs corresponding to the same spectral distance. 
\subsection{Characterization of Graph Coarsening}
\label{sec:connection}
The definitions of spectral distance are closely related to the graph coarsening framework. Specifically, the spectral distance well captures the structural changes in the graph coarsening process.
We first consider the ideal case when merged nodes within have the same normalized edge weights. 

\begin{proposition}
	Let the graph $\G_c$ be obtained by coarsening $\G$ with respect to a set of partitions $\P = \{\S_1, \S_2, \ldots, \S_n \}$. If $\P$ is selected such that every node in a partition has the same normalized edge weights,
	\begin{align}
	\label{equ:assumption1}
	    \frac{\vw(i)}{d(i)} = \frac{\vw(j)}{d(j)} \quad \text{ for all } v_i, v_j \in \S \quad \text{and} \quad \S \in \P 
	\end{align}
	then
	$$
	\SD_{\text{full}}(\G, \G_c) = 0 \quad \text{and} \quad \SD_{\text{part}}(\G, \G_c) = 0.
	$$
\end{proposition}
\begin{proof}
See the Appendix.
\end{proof}






The proposition states that under the ideal graph coarsening assumption, the spectral distance indicates that the coarse graph fully preserves the spectral properties of the original graph in the graph coarsening framework.

We next provide a more general result on how the spectral distance can capture the structural changes in the graph coarsening framework. Consider the basic coarsening where the coarse graph is formed by merging one pair of nodes (i.e. $n = N-1$), we have the following,
\begin{proposition}
	\label{prop:single_level}
	Suppose the graph $\G_c$ is coarsened from $\G$ merging a pair of nodes $v(a)$ and $v(b)$,  if the normalized edge weights of merged nodes satisfy,
	\begin{align}
	\label{eqn:connection}
	\norm{ \frac{\vw(a)}{d(a)} -  \frac{\vw(b)}{d(b)}}_1 \le \epsilon, \nonumber
	\end{align}
	then the spectral distance between the original and coarse graphs are bounded by
	$$
	\SD_{\text{full}} (\G, \G_c) \le N\epsilon \text{ and }
	\SD_{\text{part}} (\G, \G_c) \le n\epsilon.
	$$
\end{proposition}
\begin{proof}
See the Appendix.
\end{proof}

The propositions states that the spectral distance is bounded by the discrepancy of normalized edge weights of merged nodes. Thus by minimizing the nodes' edge weights within the same partitions results in better preservation of spectral properties. 

\section{ALGORITHMS}
We propose two graph coarsening algorithms to generate coarse graphs with preserved spectral properties.
The first algorithm directly follows from Proposition \ref{prop:single_level} where the coarse graphs are formed by iteratively merging graph nodes with similar normalized edge weights. The second algorithm is generalized from the spectral clustering algorithm where we leverage on the combinations of normalized Laplacian eigenvectors, combined with $k$-means clustering, to find the graph partitions and the corresponding coarse graphs. Both algorithms are shown to generate coarse graphs with small spectral distance from original graphs.
\subsection{Multilevel Graph Coarsening}
\begin{algorithm}[t!]
	\caption{Multilevel Graph Coarsening}
	\label{alg:mgc} 
	\begin{algorithmic}[1]
		\State \textbf{Input}: Graph $\G = (\V, \E, \mW)$ and target size of the coarse graph $n$.
		\State $s \gets N$
		\While{$ s > n $} 
		\For {$v_i \in \V$}
		\For {$v_j \in \mathcal{N}_i$}
		\State $d_s(i, j) = \norm{\frac{\vw(i)}{d(i)} - \frac{\vw(j)}{d(j)}}_1$
		\EndFor
		\EndFor
		\State $i_{\text{min}}, j_{\text{min}} = \arg\min_{i, j} d_s(i, j)$
		\State $s \gets s - 1$ 
		\State Merge nodes $v_{i_{\text{min}}}$ and $v_{j_{\text{min}}}$ to form the coarse graph $\G_s$.
		\EndWhile
		\State \textbf{return} $\G_n = (\V_n, \E_n, \mW_n)$
	\end{algorithmic}
\end{algorithm}
The multilevel graph coarsening algorithm follows the coarsening framework by iteratively merging pairs of nodes which share similar connections. For each iteration, the graph coarsening algorithm searches for pair of nodes with the most similar normalized edge weights and merges them into supernodes. To reduce the computational cost, we constraint the candidate pairs of graph nodes to be within 2-hop distance. We denote $\mathcal{N}_i$ as the set of nodes that are within 2-hops distance from node $v_i$. The pseudo-codes are described in Algorithm \ref{alg:mgc}. 

\paragraph{Analysis.}
The following corollary provides the spectral distance result for the multi-level graph coarsening algorithm, extended from Proposition \ref{prop:single_level},
\begin{corollary}
	\label{coro:multi_level}
	Suppose the graph $\G_c$ is coarsened from $\G$ by iteratively merging pairs of nodes $v(a_s)$ and $v(b_s)$ for $s$ from $N$ to $n+1$,  if the normalized edge weights of merged nodes satisfy,
	\begin{align}
	\norm{ \frac{\vw(a_s)}{d(a_s)} -  \frac{\vw(b_s)}{d(b_s)}}_1 \le \epsilon_s, \nonumber
	\end{align}
	then the spectral distance between the original and coarse graphs are bounded by
	\begin{align}
	\SD_{\text{full}} (\G, \G_c) \le N\sum_{s = N}^{n+1}\epsilon_s, \quad
	\SD_{\text{part}} (\G, \G_c) \le n\sum_{s = N}^{n+1}\epsilon_s \nonumber
	\end{align}
\end{corollary}
\begin{proof}
See the Appendix.
\end{proof}

\paragraph{Time complexity.}
The time complexity depends on the original graph structure as well as the coarse graph size, which can be expressed as $O(M(N+n)(N-n))$. When the coarse graph size $n$ is close to $N$ ($n\approx N$), the algorithm is efficient to generate the coarse graphs with time complexity as $O(MN)$.  However, when the coarse graphs has significantly smaller node sizes with $n\ll N$, the algorithm becomes computationally expensive with time complexity $O(MN^2)$. To address this, we propose the following spectral graph coarsening algorithm.

\subsection{Spectral Graph Coarsening} 
The spectral coarsening algorithm leverages on the combinations of eigenvectors, combined with the $k$-means clustering method, to find the coarsening partitions as well as the coarse graphs. Different from the traditional spectral clustering, we select eigenvectors with the eigenvalues corresponding to the head and tail eigenvalues as in the definition of partial spectral distance in Definition. \ref{def:partial}. Since the number of head eigenvectors $k_1$ is unknown in prior before we have the coarse graph. We iterative over possible combinations of eigenvectors, and select the coarsening results with the minimum $k$-means cost. The algorithm is described in Algorithm \ref{alg:sgc}.

\begin{algorithm}[t!]
	\caption{Spectral Graph Coarsening} 
	\label{alg:sgc} 
	\begin{algorithmic}[1]
		\State \textbf{Input}: Graph $\G = (\V, \E, \mW)$, eigenvectors $\mU$ of the normalized Laplacian $\nrL$, target size $n$.
		\State Set $k_1, k_2 \gets \arg\min_k\{i: \bm{\lambda}(k) \le 1, \bm{\lambda}(k_2 + 1) \ge 1\}$ where $k_2 \gets  N - n + k_1$
		\While{ $\bm{\lambda}(k_1) \le 1$ and  $\bm{\lambda}(k_2 + 1) \ge 1$}
		\State $\mU_{k_1} \gets [\mU(1:k_1); \mU(k_2+1:N)]$ 
		\State Apply $k$-means clustering algorithm on rows of $\mU_{k_1}$ to obtain graph partitions $\P^*_{k_1}$ that optimizes the following $k$-means cost,
		$$
		\mathcal{F}(\mU_{k_1}, \P^*_{k_1}) = \sum_{i = 1}^N \left( \vr(i) - \hspace{-2mm}\sum_{j \in \S_i} \frac{\vr(j)}{|\S_i|} \right)^2
		$$
		where $\vr(i)$ is the $i^{\text{th}}$ row of $\mU_{k_1}$.
		\State  $k_1 \gets k_1 + 1$, $k_2 = N - n + k_1$
		\EndWhile
		\State \textbf{return} coarse graph $\G_c$ generated with respect to the partitions with minimum $k$-means clustering cost as 
		$$
		\P* =\arg\min_{k_1} \mathcal{F}(\mU_{k_1}, \P^*_{k_1})
		$$
	\end{algorithmic}
\end{algorithm}


\paragraph{Analysis.}
\label{sec:sgc}		
We show that the coarse graphs obtained with Algorithm \ref{alg:sgc} optimize the partial spectral distance $\SD_{\textit{part}}$. First we assume the graph coarsening is \emph{consistent} that satisfies the following
$$
\nrL_c = \mC \nrL \mC^\top.
$$

The following theorem gives the bound of spectral distance with respective to the $k$-means costs.
\begin{theorem}
	\label{thm:sgc}
	Let the coarse graph $\G_c$ be obtained from Algorithm 2 with graph partition $\mathcal{P}^*$, and the $k$-means cost $\mathcal{F}(\mU, \mathcal{P}^*)$ satisfies $\mathcal{F}(\mU, \mathcal{P}^*) < 1$, the partial spectral distance is bounded by
	$$
	\SD_{\text{part}}(\G, \G_c) \le \frac{(n + 2)\mathcal{F}(\mU, \P^*) + 4\sqrt{\mathcal{F}(\mU, \P^*)}}{1-\mathcal{F}(\mU, \P^*)}.
	$$
\end{theorem}
\begin{proof}
See the Appendix.
\end{proof}

The theorem states that the spectral distance is bounded by the $k$-means clustering cost. When the eigenvectors have well-separated clustering structures with small $k$-means clustering cost, the spectral properties are preserved. 

\paragraph{Time complexity}
The time complexity of the spectral coarsening algorithm is $O(KTNn^2)$ where $K$ is the number of $k_1, k_2$ that satisfy $\bm{\lambda}(k_1) \le 1$ and $\bm{\lambda}(k_2+1) \ge 1$, and $T$ is the number of iterations needed until convergence.

\begin{table*}[t!]
	\caption{Classification accuracy on coarse graphs with coarsening ratio $1/5$.}
	\label{tab:result_classification}
	\begin{center}
		\begin{tabular}{ccccccc}
			{\bf Datasets}  & {\bf MUTAG} & {\bf ENZYMES} & {\bf NCI1} & {\bf NCI109} &  {\bf PROTEINS} & {\bf PTC}
			\\ \hline 
			{\bf EM} &  78.90 & 18.92  & 62.81 &61.35 &63.72& 48.56 \\
			\bf LV & 79.01 & 24.68 & 63.59 & 60.49 &62.72& 50.24\\
			\bf METIS & 77.62 & 24.79 & 59.74  & 61.64 &63.70& 49.34 \\
			\bf SC & 80.37 & 24.40 & 63.14  & 62.57 & 64.08& 50.16 \\
			\bf SGC & 80.34 & 29.19 & 63.94 & \bf 63.69&  64.70& \bf 52.76\\
			\bf MGC & \bf 81.53 & \bf 30.89 & \bf 66.07 &  63.55& \bf 65.26& 52.28 \\
			\hline 
			\bf Original & 86.58 & 37.32 & 66.39 & 64.93 & 66.60& 53.72\\
			\hline 
		\end{tabular}
	\end{center}
\end{table*}
\section{EXPERIMENTS}
In the experiments, we evaluate the graph coarsening frameworks on tasks involving real-world and synthetic graphs. 

For the first task, we compare the graph classification performance on coarse graphs coarsened from different graph coarsening methods. We evaluate the quality of graph coarsening methods by the classification performance on the coarse graphs.

The second task is on the block structure recovery of synthetic graphs sampled from the stochastic block model. We show that our graph coarsening algorithms, which optimize the spectral distance, can recover the block structures with high accuracy.

\subsection{Graph Classification with Coarse Graphs}
\label{sec:classification}
Graph classification is one of the most important graph problems with a variety of applications such as material design, drug discovery and computational neuroscience \citep{tsitsulin2018netlsd, jin2018learning, xu2018powerful,  park2013structural}. However, for some graph applications such as social network and computational neuroscience \citep{park2013structural}, graph samples usually have very large sizes, which makes the classification methods inefficient to directly apply. 

For this task, we apply the graph coarsening methods to obtain a set of coarse graphs from original graph samples. We evaluate the classification performance on coarse graphs with standard graph classification methods. The quality of graph coarsening algorithms is measured by the classification performance loss with coarse graphs compared with the original graphs. 

In the experiment, we compare with the following state-of-the art graph coarsening and partitioning algorithms,  
\begin{itemize}
	\item \textbf{Edge matching}. The coarse graphs are formed by maximum-weight matching with the weight calculated as $\mW(i, j)/\max\{d(i), d(j)\}$ \citep{dhillon2007weighted, defferrard2016convolutional}.
	\item \textbf{Neighborhood-based Local Variation}. The method is proposed to minimize the variational cost in the graph coarsening \citep{loukas2019graph, loukas2018spectrally}. 
	\item \textbf{METIS}. METIS is a standard graph partitioning algorithm based on multi-level partitioning schemes that are widely used various domains such as finite element methods and VLSI \citep{karypis1998fast}.
	\item \textbf{Spectral Clustering}. 
	Spectral clustering is a widely used graph clustering algorithm that finds densely connected graph partitions determined from the eigenvectors of the graph Laplacian  \citep{von2007tutorial}.
\end{itemize}

\subsubsection{Evaluation}
We coarsen the graph samples to graphs with node size $n = rN$ with the coarsening ratio $r$. The classification performance are evaluated with 10-fold cross validation as same as the configurations in previous works \citep{tsitsulin2018netlsd, dai2016discriminative, xu2018powerful}.

\paragraph{Datasets.} We use standard graph classification datasets for graph classification evaluation \citep{shervashidze2011weisfeiler, xu2018powerful, jin2018learning}. Each dataset contains a set of variable-sized graphs from a variety of graph applications. The specific statistics are in the Appendix. 

\paragraph{Graph Classification Method.}
We use {Network Laplacian Spectral Descriptor} (\textit{NetLSD}) combined with 1-NN classifier as the graph classification method ~\citep{tsitsulin2018netlsd}. \textit{NetLSD} was shown as an efficient graph feature extractor and achieve state-of-the art classification performance \citep{tsitsulin2018netlsd}. Note that \textit{NetLSD} extracts graph features which only depends on the graph structures without considering the node and edge features.

\begin{table*}[t!]
	\caption{Recovery Accuracy of Block Structures from Random Graphs in Stochastic Block Model}
	\label{tab:recovery}
	\begin{center}
		\begin{tabular}{cccccccc}
			{\bf p, q } & {\bf Type} & {\bf EM} & {\bf LV}& {\bf METIS} & {\bf SC}  & {\bf MGC} & {\bf SGC} \\
			\hline 			\multirow{3}{*}{\shortstack{$0.2, 0.01$}} & Associative & 0.1819 & 0.3076 & 0.7792 & \bf 0.7845 &  0.3664 & \bf 0.7845  \\
			&Dissociative & 0.0956 & 0.1071 & 0.0815 & 0.0877 & \bf 0.1093 & 0.0850  \\
			&Mixed & 0.1052 & 0.1944 &  0.2389 &0.3335 & 0.6062 & \bf 0.7107  \\
			\cline{1-8}
			\multirow{3}{*}{\shortstack{$0.5, 0.1$}} & Associative & 0.1015 & 0.1902 & 0.7820 & \bf 0.7930 & 0.2868 & \bf 0.7930  \\
			& Dissociative  & 0.0854 & 0.1068 & 0.0602 & 0.0788 & 0.1474 & \bf 0.7901  \\
			&Mixed & 0.0848 & 0.2241 & 0.2883 & 0.4074 &  \bf 0.7343 & \bf  0.7699  \\
			\cline{1-8}
			\multirow{3}{*}{\shortstack{$0.8, 0.3$}}  & Associative & 0.0823 & 0.1139 & 0.5596 & \bf 0.6532 & 0.1172 & \bf 0.6532 \\
			&Dissociative   & 0.0836 & 0.0976 & 0.0776 & 0.1342 & \bf 0.7784 & \bf 0.7931 \\
			&Mixed  & 0.0888 & 0.1503 & 0.2929 & 0.3909 & 0.5428 & \bf 0.7209  \\
			\cline{1-8}
			\hline 
		\end{tabular}
	\end{center}
\end{table*}

\subsubsection{Results}
Table \ref{tab:result_classification} shows the graph classification performance on coarse graphs with coarsening ratio 1/5. For most datasets, our graph coarsening schemes yield better classification performance compared to other graph coarsening methods. Moreover, for datasets such as NCI1, NCI109, the classification accuracy with coarse graphs achieve almost the same result as the model with the original graphs. 

\subsection{Block Recovery of Random Graph Samples}
In this experiment, we apply the coarsening algorithms to recover the block structures of random graphs sampled from stochastic block model.

The stochastic block model is a random graph model with explicit block structures, which are commonly used to evaluate graph partitioning and clustering algorithms \citep{abbe2017community, abbe2015exact}. The model is parameterized by $\mB \in {[0, 1]}^{n\times n}$ with graph nodes in blocks $i$ and $j$ are connected with probability $\mB(i, j)$. Random graph samples are generated from stochastic block model by sampling the upper triangular entries $\mW(i, j)$ following the edge probability and the lower triangular entries are constrained as $\mW(j, i) = \mW(i, j)$. 

We consider random graph models parameterized with $p$ and $q$ with the following configurations on $\mB$,

\begin{itemize}
    \item \emph{Assortative}. The diagonal entries are $p$ and off-diagonal entries are $q$.
    \item \emph{Dissortative}. The diagonal entries are $q$ and off-diagonal entries are $p$.
    \item \emph{Mixed}. The entries of $\mB$ are randomly assigned with $p$ and $q$ (each with probability 1/2).
\end{itemize}

\subsubsection{Evaluation}
We evaluate the performance of graph coarsening algorithms by measuring the discrepancy between recovered graph partitions and the ground-truth block structures. We use the \emph{Normalized Mutual Information}(\textit{NMI}) to quantify the recovery between any two graph partitions. The definition of \textit{NMI} is described in the Appendix.

For each stochastic block model setting, we use $N = 200$ and $n = 10$ with $20$ nodes for each partition. We repeat the experiment 10 times and report the average NMI results.

We compare our graph coarsening algorithms with the graph coarsening and partitioning algorithms mentioned in Section \ref{sec:classification}. Table \ref{tab:recovery} contains the average NMI results under different model settings. Our proposed methods outperform other methods under all the model configurations. In particular, our methods can achieve high recovery accuracy for dissortative and mixed settings where traditional graph partitioning algorithms fail to recover accurately. 

\section{CONCLUSION}
In this work, we propose a new framework to tackle the graph coarsening problem. We leverage on the spectral properties of normalized Laplacian matrices to define a new notion of graph distance that quantifies the differences between original and coarse graphs. We justify that the new spectral distance naturally captures the structural changes in the graph coarsening. We provide efficient graph coarsening algorithms that provably guarantee that the coarse graphs preserved spectral properties. Experiments show that our proposed methods outperform other graph coarsening algorithms on graph classification and block recovery tasks. 

\bibliographystyle{apalike}
\bibliography{ref}
\onecolumn
\appendix
\begin{center}
    \LARGE{\textbf{Supplementary Material for ``Graph Coarsening with Preserved Spectral Properties''}}
\end{center}
\section{Proof of Property 4.2, 4.3}

\begin{proof}
	\label{proof:property2}
	We start by noticing that the projection matrix $\mPi$ acts as an identity matrix w.r.t. the lifted normalized Laplacian 
    $ \nrL_l = \mPi \nrL_l \mPi$, 
    since $ \nrL_l = \C^\top \nrL_c \C = \C^\top \C \nrL_l \C^\top \C = \mPi \nrL_l \mPi$.
	Now, consider the following eigenvalue equation:
	\begin{align}
	\nrL_c\vu_c = \lambda_c \vu_c \nonumber\\
	\C\nrL_l\C^\top \vu_c = \lambda_c \vu_c \nonumber\\
	\C^\top\C \nrL_l\C^\top \vu_c = \lambda_c \C^\top \vu_c \nonumber\\
	\mPi \nrL_l \mPi \C^\top \vu_c = \lambda_c \C^\top \vu_c \nonumber\\    \nrL_l\C^\top \vu_c = \lambda_c \C^\top \vu_c \nonumber
	\end{align}
	Note that in the fourth step, we used the relation $\C^\top = \C^\top \C \C^\top = \mPi \C^\top$, which holds due to the properties of the Moore-Penrose pseudo-inverse.
	Thus, $\mC^\top \vu_c$ are eigenvectors of $\nrL_l$ with the corresponding eigenvalues of the coarse graph. 
	
	To show there are $N - n$ additional eigenvalues $1$, one can observe that $\I_N - \nrL = \mD_l^{-1/2}\mW_l\mD_l^{-1/2}$ is a rank-$n$ matrix because nodes within the same partition have exactly the same edge weights. Hence $\I_N - \nrL$ contains $N-1$ eigenvalue $0$ and correspondingly $\nrL$ contains eigenvalue $1$ with $N-n$ multiplicity.
\end{proof}

\section{Proof of Proposition 4.1, 4.2}
For the simplicity of the proof, we use the  $\rwL = \I - \mD^{-1}\mW$ to replace the original normalized Laplacian $\nrL$ to compute the Laplacian eigenvalues. Note that $\rwL$ has the same set of eigenvalues as the original normalized Laplacian $\nrL$ and the relation of the eigenvalues and eigenvectors satisfy,
$$
\rwL = \mD^{-1/2}\nrL\mD^{1/2}, \quad \vu^{\textit{rw}} = \mD^{-1/2}\vu 
$$

\subsection{Proof of Proposition 4.1}
\begin{proof}
We show that under the assumption above, the eigenvalues of the original normalized Laplacian contain the eigenvalues of coarse graph $\G_c$ plus eigenvalue $1$ with $N-n$ multiplicities.

The random-walk Laplacian of the coarse graph satisfies,
\begin{align}
    \rwL_c & = \I_n - \mD^{-1}_c\mW_c \nonumber\\
    & = \mP \I_N \mP^\mp - \mP \mD \mP^\mp \mP \mW \mP^\mp \nonumber\\
    & = \mP \I_N \mP^\mp - \mP \mD^{-1}\mW \mP^\mp \nonumber\\
    & = \mP(\I_N - \mD^{-1}\mW) \mP^\mp \nonumber\\
    & = \mP\rwL \mP^\mp \nonumber
\end{align}
%
The third equation holds because of the assumption in Equation (8). Then, the eigenvalue and eigenvector of $\rwL_c$ satisfy the following:
\begin{align}
    \rwL_c \rwu = \lambda\rwu_c \nonumber \\
    \mP\rwL \mP^\mp\rwu = \lambda\rwu_c \nonumber\\
    \mP^\mp \mP\rwL \mP^\mp\rwu = \lambda\mP^\mp\rwu_c \nonumber \\
    \rwL \mP^\mp\rwu = \lambda\mP^\mp\rwu_c \nonumber
\end{align}
that is, $\rwL$ has the eigenvalue $\lambda$ with the corresponding eigenvector $\mP^\mp\rwu$.


To see that the original graph contains $N-n$ eigenvalue 1, we consider $\mD^{-1}\mW = \I_N - \rwL$ which consists of rows of normalized edge weights with row $i$ as $\frac{\vw(i)}{d(i)}$. From the assumption in Equation (8), we have identical rows for each partition $\S_r$. Thus $\mD^{-1}\mW$ is at most rank-$n$, which indicates $\rwL$ contains $N-n$ eigenvalue 1.

Thus, the original normalized Laplacian has the same eigenvalues as the lifted graph. Both definition of spectral distances are 0.
\end{proof}
\subsection{Proof of Proposition 4.2}
\begin{proof}

	
	The normalized Laplacian of the original graph can be viewed as a perturbation of the normalized Laplacian of the lifted graph as
	$$
	\rwL = \rwL_l + \mE,
	$$
	where $\mE$ is the perturbation matrix.
	
	
	We expand the entries of $\rwL$ as follows: 
	$$
	\rwL(i, j) = \mI(i, j) - \frac{\mW(i, j)}{d(i)}.
	$$
	As the coarse graph is coarsened from merging one pair of nodes, the edge weights of the lifted graph $\G_l$ can be expressed as,
	$$
	\mW_l(i,j) = 
	\begin{cases} \frac{\mW(a, a) + \mW(a, b) + \mW(b, a) + \mW(b, b)}{4}
	 & \text{if }  i \in \{a, b\} \text{ and } j \in  \{a, b\}\\
	\frac{\mW(a, j) + \mW(b, j)}{2}    & \text{if }  i \in \{a, b\} \text{ and } j \notin  \{a, b\} \\
	\frac{\mW(i, a) + \mW(i, b)}{2}    & \text{if }  i \notin \{a, b\} \text{ and } j \in  \{a, b\} \\
	\mW(i,j) & \text{otherwise.}
	\end{cases}
	$$
	and the corresponding node degree $d_l$ is
	$$
	d_l(i) = 
	\begin{cases}
	\frac{d(a) + d(b)}{2}    & \text{if }  i \in \{a, b\} \\
	d(i) & \text{otherwise.}
	\end{cases}
	$$
	The above imply that $\rwL_l$ can be expanded as follows:
	$$
	\rwL_l = \mI(i, j) - \frac{\mW_l(i, j)}{d_l(i)} = 
	\begin{cases}
	\mI(i, j) -  \frac{\mW(a, a) +  \mW(a, b) + \mW(b, a) + \mW(b, b)}{2(d(a) + d(b))}& \text{if }  i \in \{a, b\} \text{ and } j \in  \{a, b\}\\
	\mI(i, j) -  \frac{\mW(a, j) + \mW(b, j)}{d(a) + d(b)}& \text{if }  i \in \{a, b\} \text{ and } j \notin  \{a, b\}\\
	\mI(i, j)-  \frac{\mW(i, a) + \mW(i, b)}{2d(i)}& \text{if }  i \notin \{a, b\} \text{ and } j \in  \{a, b\} \\
	\mI(i, j) - \frac{\mW(i, j)}{d(i)} & \text{otherwise} \\
	\end{cases}
	$$
	and the perturbation matrix $\mE = \rwL - \rwL_l$ is given by 
	$$
	\mE(i, j) = 
	\begin{cases}
	\frac{\mW(i, j)}{d(i)} -  \frac{\mW(a, a) + \mW(a, b) + \mW(b, a) + \mW(b, b)}{2(d(a) + d(b))}& \text{if }  i \in \{a, b\} \text{ and } j \in  \{a, b\} \\
	\frac{\mW(i, j)}{d(i)} -  \frac{\mW(a, j) + \mW(b, j)}{d(a) + d(b)}& \text{if }  i \in \{a, b\} \text{ and } j \notin  \{a, b\} \\
	\frac{\mW(i, j)}{d(i)} -  \frac{\mW(i, a) + \mW(i, b)}{2d(i)}& \text{if }  i \notin \{a, b\} \text{ and } j \in  \{a, b\} \\
	0 & \text{otherwise.} \\
	\end{cases}
	$$
	
	From \cite{weyl1912asymptotische}, we have the following bound on the eigenvalue gap between $\bm{\lambda}(i)$ and $\bm{\lambda}_l(i)$:
	$$
	|\bm{\lambda}(i) - \bm{\lambda}_l(i)| \le \norm{E}_2 
	$$
	Moreover, \cite{wolkowicz1980bounds} proved that the spectral norm $\norm{E}_2$ admits the simple upper bound:	
	$$
	\norm{\mE}_2^2  \le  \max_{i, j} \vr_i\vc_j = \max_{i}\vr_i\max_{j}\vc_j,
	$$
	
	where $\vr_i = \sum_j |\mE(i, j)|$ and $\vc_j = \sum_i |\mE(i, j)|$. 
	
	Let us focus on term $\vr_i$.
	
	Case 1: $i \notin \{a, b\}$, 
	\begin{align}
	    \vr_i &= |\frac{\mW(i, a)}{d(i)} -  \frac{\mW(i, a) + \mW(i, b)}{2d(i)}| + | \frac{\mW(i, a)}{d(i)} -  \frac{\mW(i, a) + \mW(i, b)}{2d(i)}| \nonumber \\
	    &= |\frac{\mW(i, a)}{d(i)} - \frac{\mW(i, b)}{d(i)}| \le \norm{\frac{\mW(i, a)}{d(i)} - \frac{\mW(i, b)}{d(i)}}_1 \le \epsilon \nonumber
	\end{align}
	
	Case 2: $i \in \{a, b\}$, and suppose $d(a) \le d(b)$ 
	\begin{align}
	\vr_i &= |\frac{\mW(i, a)}{d(i)} -  \frac{\mW(a, a) + \mW(a, b) + \mW(b, a) + \mW(b, b)}{2(d(a) + d(b))}| + |\frac{\mW(i, b)}{d(i)} -  \frac{\mW(a, a) + \mW(a, b) + \mW(b, a) + \mW(b, b)}{2(d(a) + d(b))}| \nonumber \\
	& + \sum_{j \notin \{a, b\}}|\frac{\mW(i, j)}{d(i)} -  \frac{\mW(a, j) + \mW(b, j)}{d(a) + d(b)}| \nonumber \\
	& \le |\frac{\mW(a, a)}{d(a)} - \frac{\mW(b, a)}{d(b)}| + |\frac{\mW(a, b)}{d(a)} - \frac{\mW(b, b)}{d(b)}| + \sum_{j \notin \{a, b\}}|\frac{\mW(a, j)}{d(a)} -  \frac{\mW(b, j)}{d(b)}| \nonumber\\
	& = \norm{\frac{\mW(i, a)}{d(i)} - \frac{\mW(i, b)}{d(i)}}_1 \le \epsilon
	\end{align}
	
	We have $\max_i \vr_i \le \epsilon$. Similarly, we can show that $\vc_j \le \epsilon$. The spectral norm of the perturbation matrix $\mE$ then is bounded by
	\begin{align}
	\norm{\mE}_2 & \le \sqrt{\max_{i} \vr_i\max_{j}\vc_j}  \le \epsilon.
	\end{align}
	Combining the above, we have the bound of each term in the spectral distance as,
	\begin{align}
	\label{eqn:term}
	|\bm{\lambda}(i) - \bm{\lambda}_l(i)| \le \epsilon
	\end{align}

	The bounds of the  full and partial spectral distance  follow the Equation \ref{eqn:term} as they contain $N$ and $n$ eigengap terms respectively.
\end{proof}

\section{Proof of Corollary 5.1}
\begin{proof}
We denote the intermediate graphs at iteration $s$ as $\G^{(s)}$ with $\G^{(N)}$ as the original graph $\G$ and $\G^{(n)}$ as the coarse graph $\G_c$. From Proposition 4.2 and the spectral distance is a distance metric over the Laplacian eigenvalues, we have the following,
\begin{align}
    \SD_{\textit{full}}(\G, \G_c) &\le \sum_{s = N}^{n+1} \SD_{\textit{full}}(\G^{(s)}, \G^{(s-1)}) \le N\sum_{s = N}^{n+1}\epsilon_s \nonumber 
\end{align}
and 
\begin{align}
    \SD_{\textit{part}}(\G, \G_c) &\le \sum_{s = N}^{n+1} \SD_{\textit{part}}(\G^{(s)}, \G^{(s-1)}) \le N\sum_{s = N}^{n+1}\epsilon_s \nonumber 
\end{align}
\end{proof}

\section{Proof of Theorem 5.2}
\begin{proof}
    We rewrite the objective of the $k$-means algorithm as the following,
$$
\mathcal{F}(\mU, \P) = \sum_{i = 1}^N \left( \vr(i) - \hspace{-2mm}\sum_{j \in \S_i} \frac{\vr(j)}{|\S_i|} \right)^2 = \| \mU - \C \C^\top\mU\|_F^2,
$$
where the matrix $\mC \in \mathbb{R}^{n \times N}$ is the normalized coarsening matrix corresponding to the graph partition $\P$. With the notation $\mPi = \mC\mC^\top$ and $\mPi^\bot = \I - \mPi$ from Section 3.2, the $k$-means objective is written as 
$$
\mathcal{F}(\mU, \P) = \|\mPi^\bot \mU\|_F^2.
$$

We express the partial spectral distance as in Definition 4.5
	\begin{align}
	    \label{eqn:sd_part}
	    \SD_{\text{part}}(\G, \G_c) = \sum_{i = 1}^{k_1}(\bm{\lambda}_c(i) - \bm{\lambda}(i)) + \sum_{j = k_2+1}^{N}(\bm{\lambda}(j) - \bm{\lambda}_c(j+n-N))
	\end{align}
	where $k_1 = \arg\max_i \{i: \bm{\lambda}_c(i) < 1\}, k_2 = N-n+k_1$.

Because of the interlacing property 4.1, we remove the absolute sign on the terms.

Correspondingly, we separate the $k$-means cost in two terms as,
	\begin{align}
	\label{eqn:kmeans}
	\mathcal{F}(\mU, \mC) = \| \mU_{k_1} - \mC\mC^\top \mU_{k_1} \|_F^2 + \| \mU'_{k_2} - \mC \mC^\top \mU'_{k_2}\|_F^2 = \|\mPi^\bot \mU_{k_1}\|_F^2 + \|\mPi^\bot \mU'_{k_2}\|_F^2\nonumber
	\end{align}
	where $\mU_{k_1}$ and $\mU'_{k_2}$  denote the eigenvectors corresponding to the smallest $k_1$ and largest $n-k_1$ eigenvalues of the original graph. We also denote $\delta_{k_1} = \|\mPi^\bot \mU_{k_1}\|_F^2$ and $\delta'_{k_2} = \|\mPi^\bot \mU'_{k_2}\|_F^2$. 
	
	We will prove the results of the two terms separately. 
	
	For the first $k_1$ eigenvalue gaps, we start by the following generalization of the Courant-Fisher theorem:
	$$
	\sum_{i \leq k_1} {\bm{\lambda}}_c(i) =  \min_{\mV^\top \mV = \mI_k} \text{tr}(\mV^\top \nrL_c \mV).
	$$
	We write $\nrL = \mS^\top \mS$ where $\mS \in \mathbb{R}^{M\times N}$ denotes the incidence matrix of the normalized Laplacian $\nrL$ with the following form
	$$
	\mS(v, e) = \begin{cases}
	\frac{1}{\sqrt{d(i)}}, \text{if } v = i \\
	-\frac{1}{\sqrt{d(j)}}, \text{if } v = j, 
	\end{cases}
	$$ 
	where $e \in \E$ with $i$ and $j$ as the connecting nodes.
	Then, the first $k_1$ eigenvalues are 
	$$
	\sum_{i \leq k_1} {\bm{\lambda}_c}(i) =  \min_{\mV^\top \mV = \mI_k} \text{tr}(\mV^\top \mC \mS^\top \mS\mC^\top\mV) = \min_{\mV^\top \mV = \mI_k} \|\mS\mC^\top\mV\|_F^2
	$$
	
	Set $\mZ = \mC\mU_{k_1}$, and suppose that $\mZ^\top \mZ$ is invertible (this will be ensured in the following). We select 
	$$
	\mV = \mZ (\mZ^\top \mZ)^{-1/2} 
	$$
	for which we have 
	$$
	\mV^\top \mV =  (\mZ^\top \mZ)^{-1/2} \mZ^\top \mZ (\mZ^\top \mZ)^{-1/2} = \mI_{k_1}
	$$
	as required.
	
	We expand the sum of eigenvalues as follows: 
	\begin{align*}
	\sum_{i \leq k_1} {\bm{\lambda}}_i 
	&= \min_{\mV^\top \mV = \mI_{k_1}} \|\mS\mC^\top\mV\|_\mF^2 \leq \|\mS\mC^\top \mZ (\mZ^\top \mZ)^{-1/2}\|_F^2  \leq \|\mS\mC^\top\mC\mU_{k_1}\|_F^2 \, \| (\mZ^\top \mZ)^{-1/2}\|_2^2.
	\end{align*}
	and use the matrix $\mPi = \C^\top\C$ and $\mPi^\bot = \mI - \mPi$ defined in Section 3.2.
	
	For the first term, we employ the triangle inequality.
	\begin{align}
	\|\mS \mC^\top\mC \mU_{k_1}\|_F^2 & = \|\mS \mPi \mU_{k_1}\|_F^2 \nonumber\\
	& = (\|\mS (\mI - \mPi^\bot) \mU_{k_1}\|_F)^2 \nonumber\\
	& \leq (\|\mS \mU_{k_1}\|_F + \|\mS \mPi^\bot \mU_{k_1}\|_F)^2 \nonumber\\
	& \leq ( \|\mS \mU_{k_1}\|_F + \|\mS \mPi^\bot\|_2 \|\mPi^\bot \mU_{k_1}\|_F)^2 
	\end{align}
	The result for  $\|\mS \mU_{k_1}\|_F$ is
	$$
	\|\mS \mU_{k_1}\|_F = 
	\sqrt{\trace{\mU_{k_1}^\top \mS^\top\mS \mU_{k_1}}} = \sqrt{\sum_{i\le k_1}\bm{\lambda}(i)}.
	$$
	On the other hand, the norm $\|\mS \mPi^\bot\|_2$ is bounded by 
	$$
	\|\mS \mPi^\bot\|_2 
	=  \sqrt{\lambda_{\text{max}}(\mPi^\bot\mS^\bot \mS \mPi^\bot)} =  \sqrt{\lambda_{\text{max}}(\nrL)} \le \sqrt{2}
	$$
	To analyze the second term, denote by $\sigma_i$ the singular values of the $k \times k$ matrix $ \mU_{k_1}^\top \mPi \mU_{k_1}$ and $\delta_{k_1} = \mathcal{F}(\mU_{k_1}, C) = \| \mPi^\bot \mU_{k_1} \|_F^2$.  The following inequality holds:
	$$
	\delta_{k_1} \geq \| \mPi^\bot \mU_{k_1} \|_2^2 = \| \mU_{k_1}^\top \mPi^\bot \mPi^\bot \mU_{k_1} \|_2 = \| \mU_{k_1}^\top \mPi^\bot \mU_{k_1} \|_2 = \| \mU_{k_1}^\top (\mI - \mPi) \mU_{k_1} \|_2 = \| \mI_k - \mU_{k_1}^\top \mPi \mU_{k_1}  \|_2  
	$$
	The inequality is equivalent to asserting that the singular values of  $\mU_{k_1}^\top \mPi \mU_{k_1}$ are concentrated around one, i.e., 
	$$
	1 - \delta_{k_1} \leq \sigma_i \leq 1 + \delta_{k_1} ~~ \text{for all} ~~ i \leq k_1.
	$$
	It follows that the smallest eigenvalue of the PSD matrix $\mZ^\top \mZ$ is bounded by
	\begin{align*}
	\bm{\lambda}_{1}(\mZ^\top \mZ) 
	&= \min_{\|\vx\|_2 = 1} x^\top \mU_{k_1}^\top \mC^\top \mC \mU_{k_1} \vx \\
	&= \min_{\vx\in \text{span}(\mU_{k_1}),\ \|\vx\|_2 = 1} x^\top \mC^\top\mC x \\
	&= \min_{\vx\in \text{span}(\mU_{k_1}),\ \|\vx\|_2 = 1}  \vx^\top \mPi x  \\
	& \geq 1-\delta_{k_1}
	\end{align*}
	We deduce that the matrix is invertible when $\delta_{k_1} < 1$ and $\mC$ is full row-rank.
	In addition, we have
	$$
	\| (\mZ^\top \mZ)^{-1/2}\|_2^2 
	=  \| (\mZ^\top \mZ)^{-1}\|_2 
	\leq \frac{1}{1 - \delta_{k_1}}.
	$$
	Putting the bounds together, gives
	$$
	\label{eqn:k1}
	\sum_{i \leq k_1} \bm{\lambda}_c(i) 
	\leq \frac{\left( \sqrt{\sum_{i \leq k} \bm{\lambda}(i)} + \sqrt{2\, \delta_{k_1}} \right)^2}{ 1 - \delta_{k_1}} 
	$$
	or equivalently
	\begin{align}
	\sum_{i \leq k} (\bm{\lambda}_c(i)  - \bm{\lambda}(i)) & \le \frac{\left( \sqrt{\sum_{i\le k} \bm{\lambda}(i)} + \sqrt{2\, \delta_{k_1}} \right)^2}{ 1 - \delta_{k_1}} - \sum_{i\le k_1} \bm{\lambda}(i)  = \frac{\delta_{k_1}(2 +\sum_{i\le k} \bm{\lambda}(i))  + \sqrt{8\delta_{k_1}\sum_{i\le k_1} \bm{\lambda}(i)}}{1-\delta_{k_1}} \nonumber
	\end{align}
	
	To prove the result for the second term in \eqref{eqn:sd_part}, we introduce the \emph{signless normalized Laplacian} $\signL = \mI + \mD^{-1/2}\mW\mD^{-1/2}$ to obtain the results of the second term in Equation. \ref{eqn:kmeans}. We follow the similar arguments using the signless normalized Laplacian. Note that the spectral properties of signless normalized Laplacian follow the relation:
	$$
	\tilde{\bm{\lambda}}(i) = 2 - \bm{\lambda}(N+1-i) \text{ and } \tilde{\mU}(i) = \mU(N+1-i)
	$$
	Then, the eigengaps between largest eigenvalues abide to
	\begin{align}
	\label{key}
	\sum_{j = k_2+1}^{N}(\bm{\lambda}(j) - \bm{\lambda}_c(j+n-N)) & = \sum_{j = 1}^{n-k} \bm{\lambda}(N+1-j) - \bm{\lambda}_c(n+1-j) \nonumber\\
	& = \sum_{j = 1}^{n-k} (\tilde{\bm{\lambda}}_c(j) - \tilde{\bm{\lambda}}(j)) \nonumber\\
	& \le \frac{\delta'_{k_2}(\sum_{j\le n-k_1}  2 + \tilde{\bm{\lambda}}(j)) + \sqrt{8\delta'_{k_2}\sum_{j\le n-k_1} \tilde{\bm{\lambda}}(j)}}{1-\delta'_{k_2}}. \nonumber 
	\end{align}
	
	Combining the above, we obtain the following result:
	\begin{align}
	\SD(\G, \G_c) & \le \frac{\delta_{k_1}(2 +\sum_{i\le k} \bm{\lambda}(i))  + \sqrt{8\delta_{k_1}\sum_{i\le k_1} \bm{\lambda}(i)}}{1-\delta_{k_1}} + \frac{\delta'_{k_2}(\sum_{j\le n-k_1}  2 + \tilde{\bm{\lambda}}(j)) + \sqrt{8\delta'_{k_2}\sum_{j\le n-k_1} \tilde{\bm{\lambda}}(j)}}{1-\delta'_{k_2}} \nonumber\\
	& \le \frac{(n + 2)\mathcal{F}(\mU, \mC) + 4\sqrt{\mathcal{F}(\mU, \mC)}}{1-\mathcal{F}(\mU, \mC)} \nonumber
	\end{align}
	In the last step, we use the following bounds:
	$$
	\delta_{k_1} \le \mathcal{F}(\mU, \mC),
	\delta'_{k_2} \le \mathcal{F}(\mU, \mC),
	$$
	$$
	\sum_{i\le k_1} \bm{\lambda}(i) \le k_1, \sum_{j\le n-k_1} \tilde{\bm{\lambda}}(j) \le n-k_1
	$$
	$$
	\sqrt{k_1} + \sqrt{n-k_1} \le \sqrt{2n}.
	$$
\end{proof}
\section{Additional Material for Experiments}
\subsection{Graph Classification Dataset}
The statistics of the graph classification benchmarks are in Table \ref{tab:graph_stat}.
\begin{table*}[h!]
	\caption{Statistics of the graph benchmark datasets.}
	\label{tab:graph_stat}
	\begin{center}
		\begin{tabular}{ccccccc}
			{\bf Datasets}  & {\bf MUTAG} & {\bf ENZYMES} & {\bf NCI1} & {\bf NCI109} &  {\bf PROTEINS} & {\bf PTC}
			\\ \hline 
			\bf Sample size & 188 & 600 & 4110 &4127 & 1108 & 344 \\
			\bf Average $|V|$ & 17.93 & 32.63  & 29.87 & 29.68 & 39.06 & 14.29\\
			\bf Average $|E|$& 19.79 & 62.14 & 32.3 & 32.13 & 72.70 & 14.69\\
			\bf 	\# classes  & 2 & 6 & 2  &2 & 2 & 2 \\
			\hline 
		\end{tabular}
	\end{center}
\end{table*}
\subsection{Definition of Normalized Mutual Information}
We denote $C_1$ and $C_2$ are two where $C(i)$ represents the set of nodes with label $i$. We define the NMI as,
$$
NMI(C_1, C_2) = \frac{MI(C_1, C_2) }{\frac{1}{2}(H(C_1) + H(C_2))}
$$
where $MI (C_1, C_2)$ is the mutual information defined as,

$$
MI(C_1, C_2) = \sum_{i = 1}^{n}\sum_{j = 1}^{n} p ( C_1(i) \cap C_2(j)) \log \left( \frac{p\left( C_1(i)\cap C_2(j)\right)}{p\left( C_1(i) \right) p\left(C_2(j) \right)}\right)
$$

$H(C)$ is the entropy defined as,
$$
H(C) = -\sum_{i = 1}^{n}p(C(i)) \log p(C(i))
$$
The probability $p(C(i))$ is approximated as the ratio of partition $i$ as $p(C(i)) = \frac{|C(i)|}{N}$.

\end{document}

%% file: math_commands.tex
\usepackage{amsmath,amsfonts,bm}

\def\eqref#1{equation~\ref{#1}}

\def\1{\bm{1}}

\def\vc{{\bm{c}}}

\def\vr{{\bm{r}}}

\def\vu{{\bm{u}}}

\def\vw{{\bm{w}}}
\def\vx{{\bm{x}}}

\def\mB{{\bm{B}}}
\def\mC{{\bm{C}}}
\def\mD{{\bm{D}}}
\def\mE{{\bm{E}}}
\def\mF{{\bm{F}}}

\def\mI{{\bm{I}}}

\def\mP{{\bm{P}}}

\def\mS{{\bm{S}}}

\def\mU{{\bm{U}}}
\def\mV{{\bm{V}}}
\def\mW{{\bm{W}}}

\def\mZ{{\bm{Z}}}

\DeclareMathAlphabet{\mathsfit}{\encodingdefault}{\sfdefault}{m}{sl}
\SetMathAlphabet{\mathsfit}{bold}{\encodingdefault}{\sfdefault}{bx}{n}

\newcommand{\E}{\mathbb{E}}

